\documentclass[letterpaper, 10 pt, conference]{ieeeconf}  % Comment this line out if you need a4paper

\IEEEoverridecommandlockouts                              % This command is only needed if 
\usepackage{epsfig} % for postscript graphics files
\usepackage{mathptmx} % assumes new font selection scheme installed
\usepackage{times} % assumes new font selection scheme installed
\usepackage{amsmath}
\usepackage{multirow} 
\usepackage{subfig}
\usepackage{amssymb}
\usepackage{algorithm}
\usepackage{hyperref}
\usepackage{algpseudocode}
\usepackage{booktabs} 
\usepackage{array}
\usepackage{textcomp}
\usepackage{url}
\usepackage{verbatim}
\usepackage{graphicx}
\usepackage{cite}
\usepackage{makecell}
\usepackage{caption}
\usepackage{xcolor}
\usepackage{tcolorbox}
\tcbuselibrary{breakable}

\title{\textbf{N}atural Gr\textbf{a}dient Gaussia\textbf{n} Appr\textbf{o}ximation Filter with \\ Positive Definiteness Guarantee}

\author{Tianyi Zhang$^{1}$, Wenhan Cao$^{1}$, and Shengbo Eben Li$^{*}$ 
\thanks{This study is supported by Tsinghua-Efort Joint Research Center for EAI Computation and Perception.}
\thanks{$^{1}$Tianyi Zhang and Wenhan Cao are with the School of Vehicle and Mobility, Tsinghua University, Beijing, China (E-mail: \{zhangtia24, cwh19 \}@mails.tsinghua.edu.cn).}%
\thanks{$^{*}$Shengbo Eben Li is with the School of Vehicle and Mobility and College of Artificial Intelligence, Tsinghua University, Beijing, China (E-mail: lish04@gmail.com).}
\thanks{Corresponding Author: Shengbo Eben Li.}
}
\begin{document}

\maketitle
\thispagestyle{empty}
\pagestyle{empty}

%%%%%%%%%%%%%%%%%%%%%%%%%%%%%%%%%%%%%%%%%%%%%%%%%%%%%%%%%%%%%%%%%%%%%%%%%%%%%%%%
\begin{abstract}
Popular Bayes filters often apply linearization techniques, such as Taylor expansion or stochastic linear regression, to enable the use of the Kalman filter structure, but this can lead to large errors in strongly nonlinear systems.
The recently proposed NANO filter addresses this issue by interpreting the prediction and update steps of Bayesian filtering as two distinct optimization problems and solving them through moment matching and natural gradient descent, thereby avoiding model linearization errors.
However, the natural gradient update in NANO can occasionally diverge because the posterior covariance in its iteration may lose positive definiteness. Our analysis shows that the posterior covariance is the sum of the inverse prior covariance and the expected Hessian of the log-likelihood function, and that the indefiniteness of the latter term is the root cause of update failure.
To address this issue, we propose two remedies. The first approximates the log-likelihood Hessian using the Gauss–Newton method, representing it as the self-adjoint product of the Jacobian of the normalized measurement residual, which is guaranteed to be positive semi-definite. The second reformulates the covariance update as an exponential-form update of the Cholesky factor and reconstructs the covariance via its Gram matrix, which ensures positive definiteness.
Experiments on three classical nonlinear systems demonstrate that the proposed NANO filter with guaranteed positive definiteness outperforms popular members of the Kalman filter family and original NANO filter.
\end{abstract}

\section{Introduction}
State estimation plays a central role in many industrial control applications such as robotics, manufacturing, aerospace, and transportation \cite{grewal2010applications,liu2024convolutional,zhang2025robust}. In these fields, it is essential to accurately infer the internal state of a dynamical system based on noisy sensor measurements and imperfect model. The measurements often contain noise with varying statistical properties. The model used to describe system behavior are inherently imprecise, leading to uncertainty. The core objective of a filtering algorithm is to balance the uncertainties introduced by both sensor noise and model inaccuracies.

Bayesian filtering provides a fundamental framework for state estimation by sequentially updating the probability distribution of the system state \cite{sarkka2023bayesian}. When system dynamics and measurement models are linear with Gaussian noises, this framework simplifies to the Kalman filter (KF), which offers elegant analytical expressions for the estimated mean and covariance, making it computationally efficient \cite{kalman1960new}. However, real-world systems are often nonlinear, where the standard KF cannot be directly applied due to its linear assumptions.
% Alternatives like particle filters and moving horizon estimation address nonlinearities by approximating arbitrary distributions or optimizing over sliding windows, respectively \cite{del1997nonlinear,liu1998sequential,cao2025robust}. Yet, these methods usually demand high computational resources, limiting their use in real-time industrial applications. As a more efficient compromise significantly reducing computational cost while still accommodating some nonlinear effects 
To address nonlinearities, Gaussian filtering is a primary method that balances estimation speed and accuracy. It approximates nonlinear functions to maintain the Gaussian assumption for state estimates, enabling Kalman-like updates \cite{thrun2002probabilistic}. The extended Kalman filter (EKF) and iterated EKF (IEKF) rely on first-order Taylor expansions to linearize around operating points \cite{smith1962application,bell1993iterated}. Other variants, such as the unscented Kalman filter (UKF), Gaussian-Hermite Kalman filter, and cubature Kalman filter, use stochastic regression techniques to better capture the transformed mean and covariance \cite{julier1995new,julier2004unscented,arasaratnam2007discrete,arasaratnam2009cubature}. The posterior linearization filter (PLF) further improves accuracy via iterative regression during updates \cite{garcia2015posterior}. These Gaussian filters are widely favored in nearly linear settings for real-time use \cite{sarkka2023bayesian}, but in highly nonlinear environments, they can still produce substantial estimation errors.

Motivated by these challenges, a recent important breakthrough in Gaussian filtering is the Natural Gradient Gaussian Approximation (NANO) filter \cite{cao2024nonlinear}, which reformulates the prediction and update steps of Bayesian filtering as two distinct parameter optimization problems over the Gaussian mean and covariance.
In the prediction step, the NANO filter applies moment matching to determine the mean and covariance of the prior distribution. In the update step, it employs natural gradient descent to directly optimize the posterior parameters, thereby avoiding linearization errors.
The NANO filter significantly improves estimation accuracy in highly nonlinear systems while maintaining computational simplicity. However, the natural gradient iterations in NANO filter can occasionally diverge because the covariance update may lose positive definiteness, particularly in scenarios with strong nonlinearity or poor initial conditions. While the original NANO filter uses EKF initialization to stabilize iterations, this strategy is flawed in EKF-challenged systems (e.g., severe nonlinearity), where suboptimal initialization points degrade performance.

In this paper, we address these limitations by proposing two different approaches to guarantee the positive definiteness of the covariance update in the NANO filter. Specifically, our contributions are summarized as follows:
\begin{itemize}
    \item We identify that the loss of positive definiteness in NANO’s covariance update arises from the indefiniteness of the log-likelihood Hessian. The covariance update formula can be decomposed into the sum of two terms: (i) the inverse of the prior covariance, which is strictly positive definite, and (ii) the log-likelihood Hessian, consisting of a Fisher information term and a residual-induced curvature term. While the Fisher information term is always positive semidefinite, the residual-induced curvature term introduces negative curvature, making the overall Hessian indefinite. As a result, the updated covariance is not guaranteed to remain positive definite, potentially leading to divergence.
    \item We propose two complementary strategies to ensure the covariance remains positive definite throughout the NANO filter's iterations. The first method employs a Gauss-Newton approximation of the log-likelihood Hessian, reformulating it as the self-adjoint product of the Jacobian of the normalized measurement residual. This representation inherently guarantees positive semi-definiteness while preserving essential curvature information. The second method reformulates the covariance update as an exponential-form update applied to the Cholesky factor, then reconstructs the covariance matrix through its Gram matrix representation, ensuring positive definiteness by reconstruction.
\end{itemize}

\section{Algorithm Overview}\label{sec.2}
This section discusses an enhanced algorithmic design of the NANO filter, whose principle was first proposed in \cite{cao2024nonlinear}. It has been proved that NANO can significantly improve the filter's performance to nonlinear dynamics, as it formulates Bayesian filtering as two distinct optimization problems and solves them directly, thereby avoiding the model linearization errors. 
According to  \cite{cao2024nonlinear}, NANO has two versions. One is derivative-free, which leverages Stein's lemma to bypass these derivative calculations. Interested readers can see \cite{cao2025algorithm} for its algorithm design. The other is derivative-based, which necessitates computing the first and second derivatives of the log-likelihood function during the update iteration. This article will focus on the algorithm design of derivative-based NANO, with emphasis on how to avoid the covariance update losing positive definiteness.
% provides clearer insight into the root cause of the covariance update losing positive definiteness.
% deriavative-free 的版本

Considering a nonlinear dynamic system:
\begin{equation}\label{eq.SSM}
\begin{aligned}
x_{t+1} &= f(x_t, u_t) + \xi_t,
\\
y_t &= g(x_t) + \zeta_t, 
\end{aligned}
\end{equation}
where $x_t \in \mathbb{R}^n$ is the system state, $u_t \in \mathbb{R}^l$ is the control input, $y_t \in \mathbb{R}^m$ is the noisy measurement. The function ${f}: \mathbb{R}^n \to \mathbb{R}^n$ is the transition function, and ${g}: \mathbb{R}^n \to \mathbb{R}^m$ is the measurement function; ${\xi}_t$ denotes a zero-mean  process noise with covariance $Q \in \mathbb{R}_{+}^{n \times n}$, and ${\zeta}_t$ denotes a zero-mean measurement noise with covariance $R \in \mathbb{R}_{++}^{m \times m}$. 

The NANO filter still follows the two-step structure of Bayesian filtering: prediction and update, which means it maintains a prior $p(x_t|y_{1:t-1})$ and a posterior $p(x_t|y_{1:t})$ at each time instant. As a class of Gaussian filters, NANO approximates both of them as Gaussian distributions:
\begin{equation}\label{eq.Gaussian approximation}\nonumber
\begin{aligned}
p(x_t|y_{1:t-1}) &\approx \mathcal{N}(\hat{x}_{t|t-1}, P_{t|t-1}), \\ p(x_t|y_{1:t}) &\approx \mathcal{N}(\hat{x}_{t|t}, P_{t|t}).   
\end{aligned}    
\end{equation}
Here, $\hat{x}_{t|t-1} \in \mathbb{R}^n$ and $ P_{t|t-1} \in \mathbb{R}^{n\times n}$ are called prior mean and prior covariance, and $\hat{x}_{t|t} \in \mathbb{R}^n$ and $ P_{t|t}\in \mathbb{R}^{n\times n}$ are called  posterior mean and poseterior covariance. 
In the prediction step, the first two moments of the prior distribution are calculated based on the transition model and control input. In the update step, the natural gradient descent is employed to solve for the optimal Gaussian posterior based on the measurement model and a new measurement.

% The prediction step calculates the prior mean and prior covariance based on the transition model. The update step calculates the optimal Gaussian posterior based on the measurement model and a new measurement.

The moment matching (MM) technique is an important computational tool used in both steps of the NANO filter. It approximates the mean \(\mu^\prime\) and covariance \(\Sigma^\prime\) of a random variable \(x \sim \mathcal{N}(\mu,\Sigma)\) after passing through a nonlinear function \(f(\cdot)\), which is concisely expressed as $\{\mu', \Sigma'\} = \mathrm{MM}(\mu, \Sigma; f(\cdot))$.
% \begin{equation}
% \nonumber
% \{\mu', \Sigma'\} = \mathrm{MM}(\mu, \Sigma; f(\cdot)).
% \end{equation}
Its general computational framework is
\begin{equation}
\begin{aligned}
\nonumber
\mu^\prime &= \sum_{i=0}^{N-1} w_i\,f(\chi_i),\\
\Sigma^\prime &= \sum_{i=0}^{N-1} w_i\!\left[f(\chi_i)-\mu^\prime\right]\!\left[f(\chi_i)-\mu^\prime\right]^{\top}, 
\end{aligned}
\end{equation}
where \(\chi_i\) is the collocation point and \(w_i\) is the weighting coefficent. Depending on how collocation point and weighting coefficent are constructed, MM can be implemented in different forms, such as unscented transformation \cite{julier1995new,julier2004unscented}, Gauss–Hermite quadrature \cite{arasaratnam2007discrete} or spherical–radial cubature \cite{arasaratnam2009cubature}, etc. 

Specifically, the derivative-based version of NANO filter is summarized as follows: at each time instant $t$, first record the current-time signals and then perform prediction step and update step; repeat this process until filtering is terminated. Before starting NANO, we need to initialize the posterior mean \(\hat{x}_{0|0}\) and posterior covariance \(P_{0|0}\) of the first time instant \(t=0\). The threthold \(\gamma\) is the only parameter need to be specified in NANO. Starting from \(t=1\), NANO repeats the following process:
\begin{tcolorbox}[
  colframe=black,        % 边框颜
  colback=white,         % 背景颜色
  boxrule=1pt,           % 边框宽度
  arc=0pt,               % 直角边框（0pt为直角，非0为圆角）
  left=10pt, right=10pt, % 左右边距
  top=5pt, bottom=5pt,    % 上下边距
  breakable
]
\textbf{(1) Data Collection}

Record control input \(u_{t-1}\) and measurement \(y_t\).
\\

\textbf{(2) Prediction Step}

(2-1) Calculate prior mean:
\begin{equation}\label{eq.prior mean}
\left\{ \hat{x}_{t|t-1}, W  \right\}= \mathrm{MM}(\hat{x}_{t-1|t-1}, P_{t-1|t-1}; f(\cdot, u_{t-1})).
\end{equation}

(2-2) Calculate prior covariance:
\begin{equation}\label{eq.prior covariance}
P_{t|t-1} = W+Q.    
\end{equation}
\vspace{0.05cm}

\textbf{(3) Update Step}

(2-1) At each time instant $t$, initialize posterior mean $\hat{x}_t^{(0)}=\hat{x}_{t|t-1}$ and posterior covariance $P_t^{(0)}=P_{t|t-1}$. Here, $P_t^{(0)}$ is required to be positive definite.  

(2-2) For $k$, do iteration:

\quad \quad (i) Define the log-likelihood function:
\begin{equation}
\begin{aligned}
\label{eq.obj_func}
\ell(x,y_t) &= \frac{1}{2}(y_t - g(x))^\top R^{-1}(y_t-g(x)),
\end{aligned}    
\end{equation}

\quad \quad (ii) Calculate expected coefficients:
\begin{equation}
\begin{aligned}
\left\{ V_x^{(k)}, \cdot  \right\} &= \mathrm{MM}(\hat{x}_{t}^{(k)}, P_{t}^{(k)}; \frac{\partial \ell(x, y_t)}{\partial x}),
\\
\left\{ V_{xx}^{(k)}, \cdot  \right\}&= \mathrm{MM}(\hat{x}_{t}^{(k)}, P_{t}^{(k)}; \frac{\partial^2 \ell(x, y_t)}{\partial x^2} )).
\end{aligned}
\end{equation}

\quad \quad (iii) Calculate intermediate mean and covariance:
\begin{subequations}
\begin{align}
\label{eq.inter_cov}
\left( P_t^{-1} \right)^{(k+1)} =& P_{t|t-1}^{-1} + V_{xx}^{(k)},\\
\label{eq.inter_mean}
\hat{x}^{(k+1)}_t =& \hat{x}^{(k)}_t - P_t^{(i+1)} 
V_x^{(k)} \nonumber \\
&-
P_t^{(k+1)} P_{t|t-1}^{-1}(\hat{x}_t^{(k)} - \hat{x}_{t|t-1}).
\end{align}
\end{subequations}

\quad \quad (iv) Stopping criterion: 
\begin{equation}\label{eq.stopping criterion}
\nonumber
\|P^{(k+1)}_t - P^{(k)}_t \|
< \gamma.
\end{equation}

(2-3) Update posterior mean and posterior covariance
\begin{equation}
\nonumber
\hat{x}_{t|t} = \hat{x}_t^{(k+1)}, \quad
P_{t|t} = P_t^{(k+1)}. 
\end{equation}
\end{tcolorbox}

%解释一下自然梯度的结果
Note that \eqref{eq.inter_cov} and \eqref{eq.inter_mean} come from natural gradient descent of minimizing the posterior cost with respect to the Gaussian parameters. By leveraging the Fisher information matrix that captures the curvature of the parameter space, the intermediate mean and covariance are adjusted to perform the steepest descent on the Gaussian manifold.

\section{Methods for Ensuring Positive Definiteness}

In the last section, we explained how the NANO filter uses natural gradient iterations to compute the posterior estimate, effectively avoiding linearization errors. However, during the iterative update process in \eqref{eq.inter_cov}, the covariance matrix is not guaranteed to remain positive definite. This stems from the structure of the covariance update rule: while the prior covariance \(P_{t|t-1}\) is strictly positive definite, the Hessian term involved in the update introduces indefinite components. Specifically, the log-likelihood Hessian takes the form
\begin{equation}
\begin{aligned}
\label{eq.hessian}
% \frac{\partial \ell(x_t, y_t)}{\partial x_t} &= -\left.\frac{\partial h}{\partial x}\right|_{x=x_t}^\top R^{-1}(y_t - h(x_t)) \\
\frac{\partial^2 \ell(x, y_t)}{\partial x^2} = G^\top R^{-1}G - \frac{\partial^2 g}{\partial x^2}^\top R^{-1}(y_t - g(x)),
\end{aligned}
\end{equation}
where \(G = {\partial g}/{\partial x}\) is the Jacobian of the measurement function. The first term \(G^\top R^{-1} G\) is always positive semi-definite. In contrast, the second term depends on the measurement residual \((y_t - g(x))\) and the Hessian of the measurement function \({\partial^2 g}/{\partial x^2}\), which can take either positive or negative values. As a result, the full Hessian is generally indefinite. Therefore, using the iteration in \eqref{eq.inter_cov} may cause the covariance matrix to become non-positive definite at certain time instant, leading to the divergence of the NANO filter. To address this problem, we introduce two methods that guarantee the positive definiteness of the covariance updates.

\subsection{Hessian Approximation}
The first approach is to directly approximate the log-likelihood Hessian \eqref{eq.hessian} as a positive semi-definite matrix. We define the normalized measurement residual as \(r(x,y_t)= R^{-1/2}(y_t - g(x))\). With this representation, the objective function \eqref{eq.obj_func} can be transformed into a least-squares form as \(\ell(x,y_t) = \frac{1}{2}\|r(x,y_t)\|_2^2 \). Therefore, we have \(\ell(x+\Delta x,y_t) = \frac{1}{2}\|r(x+\Delta x,y_t)\|_2^2 \), and then performing a Taylor expansion on both sides of the equation, we have
\begin{equation}
\begin{aligned}
\ell(x+\Delta x,y_t) \approx& \ell(x,y_t) + \frac{\partial \ell}{\partial x}^\top\Delta x + \frac{1}{2}\Delta x^\top\frac{\partial^2 \ell}{\partial x^2}\Delta x, \\
\frac{1}{2}\|r(x+\Delta x,y_t)\|_2^2 \approx& \frac{1}{2}\|r(x,y_t) + J(x)\Delta x\|_2^2 \\
=& \frac{1}{2}\|r(x,y_t)\|_2^2 + r(x,y_t)^\top J(x)\Delta x \\
& + \frac{1}{2}\Delta x^\top J(x)^\top J(x)\Delta x,
\end{aligned}
\end{equation}
where \(J(x) = {\partial r}/{\partial x} = -R^{-1/2}G\) denote the Jacobian of \(r(x,y_t)\). Meanwhile, it is easy to verify that \({\partial \ell^\top}/{\partial x} =r(x,y_t)^\top J(x) \), so in the second-order Taylor sense, the log-likelihood Hessian can be approximated by the self-adjoint product of the residual’s Jacobian as
\begin{equation}
\frac{\partial^2 \ell(x, y_t)}{\partial x^2} \approx J(x)^\top J(x)=G^\top  R^{-1}G.
\end{equation}

This approximation ensures that the Hessian is positive semidefinite, which in turn guarantees that the covariance matrix remains positive definite throughout the entire iterative process. The proof is as follows:

\begin{proof}
For any non-zero vector \(x \in \mathbb{R}^n\), consider the quadratic form as
\begin{equation}
\begin{aligned}
\ x^\top\frac{\partial^2 \ell(x, y_t)}{\partial x_t^2}x&\approx x^\top G^\top R^{-1} Gx, \\
&=(Gx)^\top R^{-1} (Gx).
\end{aligned}
\end{equation}
Since \(R\) is positive definite, \(R^{-1}\) is also positive definite. Thus, \((G x)^\top R^{-1} (G x) \geq 0\) for all x, with equality if and only if \(G x = 0\). This confirms that the approximated Hessian is positive semi-definite. Moreover, since the MM step in \eqref{eq.inter_cov} involves summation and \(P_{t|t-1}^{-1}\) is strictly positive definite, the iteratively obtained \((P_t^{-1})^{(k+1)}\) is guaranteed to be positive definite.
\end{proof}

Note that this approximation of the Hessian actually ignores the second term of the exact Hessian \eqref{eq.hessian} and is equivalent to the Gauss–Newton method \cite{bell1993iterated}. This is reasonable because, when the normalized residual is relatively small, such as during stable iterations where the Gaussian approximation closely matches the true posterior, the second term’s contribution to the Hessian becomes insignificant compared to the dominant positive semi-definite first term.

\subsection{Cholesky Decomposition}
Another approach exploits the fact that any positive-definite matrix admits a Cholesky decomposition.  
Before the iteration of update step at time instant $t$ begins, the inverse covariance matrix is factorized as \((P_t^{-1})^{(k)} = (\Lambda_t^{(k)})(\Lambda_t^{(k)})^\top\), where \(\Lambda_t^{(k)}\) is a lower-triangular matrix. Based on the conclusions in \cite{lin2021tractable}, the original covariance iteration \eqref{eq.inter_cov} can be directly written as an iteration for the decomposed matrix \(\Lambda_t^{(k)}\) as
\begin{equation}
\label{eq.inter_cov_2}
\Lambda^{(k+1)} = \Lambda^{(k)}\exp_m(\frac{1}{2}(\Lambda^{(k)})^{-1}(V_{xx}^{(k)}+P_{t|t-1}^{-1})(\Lambda^{(k)})^{-\top}),
\end{equation}
where \(\exp_m(\cdot)\) is the matrix exponential function. To make the computation more tractable, we can further simplify \eqref{eq.inter_cov_2} by using a first-order approximation of the matrix exponential map, yields
\begin{equation}
\label{eq.inter_cov_chole}
\Lambda^{(k+1)} \approx \Lambda^{(k)} +\frac{1}{2}(V_{xx}^{(k)}+P_{t|t-1}^{-1})(\Lambda^{(k)})^{-\top}.
\end{equation}
During the iteration, we update \(\Lambda_t\) directly and reconstruct the inverse of covariance using \((P_t^{-1})^{(k+1)} =(\Lambda_t^{(k+1)})(\Lambda_t^{(k+1)})^\top+\epsilon I \), where \(\epsilon>0\) is a small tunable parameter. This factorization and reconstruction can ensure that the covariance matrix remains positive definite throughout. The proof is as follows:

\begin{proof}
For any non-zero vector \(x \in \mathbb{R}^n\), we have
\begin{equation}
\nonumber
\begin{aligned}
x^\top(\Lambda_t^{(k+1)})(\Lambda_t^{(k+1)})^\top x=\|(\Lambda_t^{(k+1)})^\top x\|^2_2 \geq 0,
\end{aligned}
\end{equation}  
so the matrix \((\Lambda_t^{(k)}) (\Lambda_t^{(k)})^\top\) is positive semi-definite. Furthermore, by adding a small positive definite matrix \(\epsilon I\), the inverse of the covariance \((P_t^{-1})^{(k+1)}\) becomes positive definite, which means the covariance \(P_t^{(k+1)}\) is positive definite.
\end{proof}

In summary, both approaches ensure that the covariance matrix remains positive definite throughout the natural gradient updates. However, the second approach involves performing Cholesky decomposition and first-order approximation at each iteration, which results in higher computational complexity and potential errors. On the other hand, the first approach offers second-order approximation accuracy and avoids redundant calculations. Therefore, we use the first approach as the primary method for the standard NANO filter.

\section{Experimental Results}
\label{sec.experiments}
In this section, we evaluate our proposed method, the NANO filter with Hessian approximation that guarantees positive definiteness, across three benchmark nonlinear systems to demonstrate its applicability over classical filtering algorithms under both ideal and model mismatch conditions. We conduct $N = 100$ Monte Carlo (MC) experiments for each system, to avoid the influence of randomness. In each MC experiment, the chosen evaluation metric is the root mean square error (RMSE), defined as \(\text { RMSE }=\sqrt{\frac{1}{M\cdot n} \sum_{t=1}^M\left\|x_t-\hat{x}_t\right\|^2} \).
Here, $x_t, \hat{x}_t$ stand for the real and estimated state, $M$ is the length of running steps and $n$ is the dimension of state. As it is impossible to compare with all existing Gaussian filters, we focus on the most popular ones as baselines for our analysis, including the EKF, UKF, IEKF, and PLF. 

\subsection{FM Demodulator}
We first consider the discrete-time non-linear system of FM demodulator \cite{ singh2024inverse}. The state space model is given by
\begin{equation}
\begin{aligned}
x_{t+1} &= \left[\begin{array}{cc}
\exp (-T / \beta) & 0 \\
-\beta \exp (-T / \beta)-1 & 0.1
\end{array}\right]x_t+ \xi_t, \\
y_t &= \sqrt{2} \begin{bmatrix}
\sin{\theta_t}\\
\cos{\theta_t}
\end{bmatrix} + \zeta_t.
\end{aligned}
\end{equation}
Here, $\xi_t \sim \mathcal{N}(\mathbf{0},\begin{bmatrix}
0.01 & 0\\
0 & 1
\end{bmatrix}), \zeta_t \sim \mathcal{N}\left(\mathbf{0}, \mathbf{I}_2\right)$, $T=2 \pi / 16$ and $\beta=100$. In addition to the ideal conditions, we also consider two common types of model mismatch to verify the robustness of the NANO filter: system modeling error and measurement outlier. The former is simulated by applying perturbations of varying percentages to the parameters, while the later is simulated by introducing low-probability, high-variance noise to the measurements.
\begin{itemize}
    \item System modeling error: \(\beta \leftarrow \beta (1 + o), o\in \{0, \pm1\%, \pm5\%,\pm10\%\}\),
    \item Measurement outlier: \(\zeta_t \sim (1-k)\mathcal{N}(0,   \cdot\mathbf{I}_{2 \times 2})+ k\mathcal{N}(0,  100 \cdot\mathbf{I}_{2 \times 2})\), \(k\in \{0, 0.01, 0.04,0.07,0.1\}\).
\end{itemize}

\begin{figure}[!t]
\centering
\includegraphics[width=0.4\textwidth]{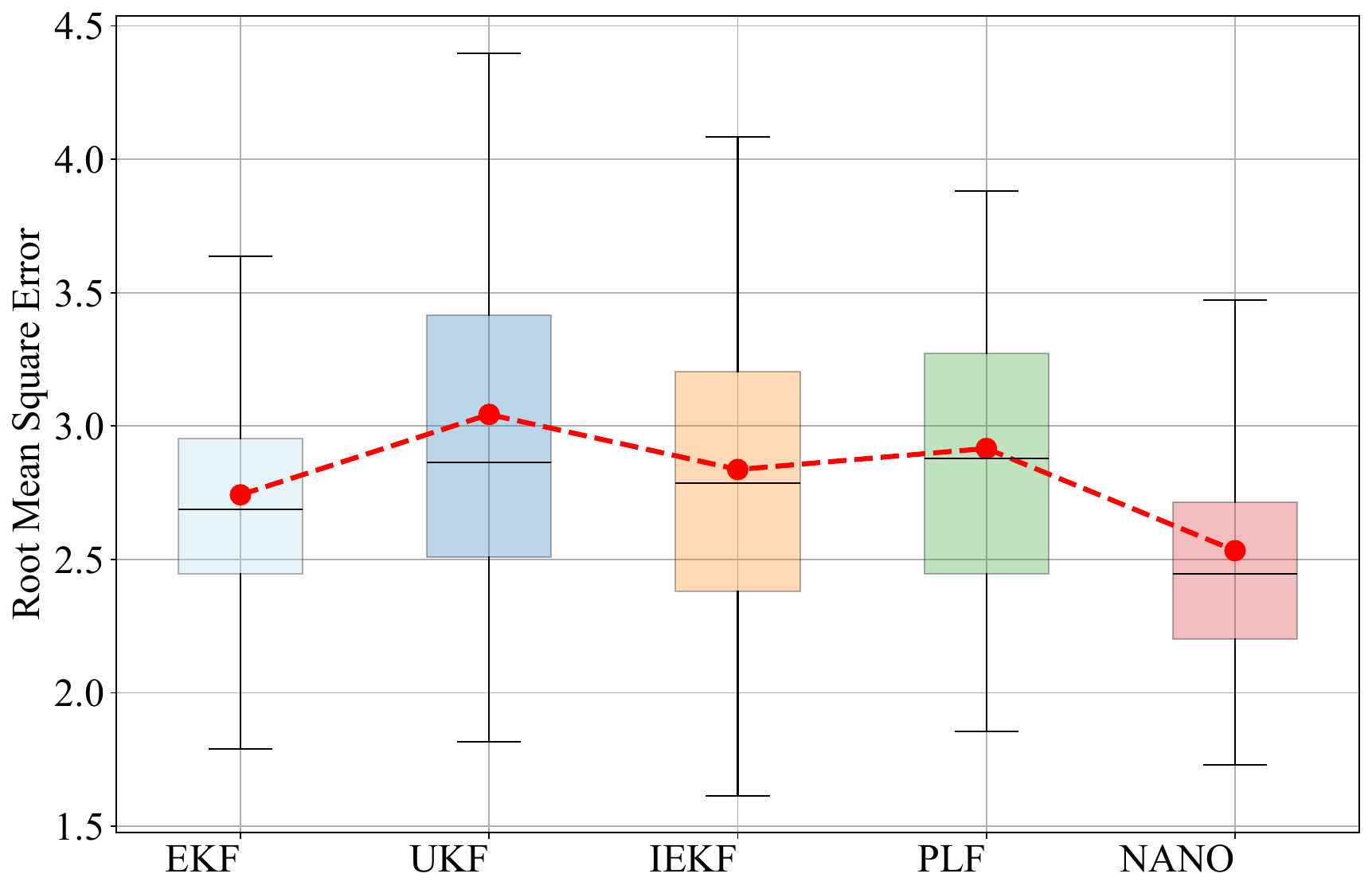}

\caption{Box plot of RMSE over all MC experiments for FM Demodulator. Note that red point `` $\textcolor{red}{\bullet}$
 " represents the average RMSE.}
\label{fig.Demodulator}
\end{figure}

\begin{figure}[!t]
    \centering
    \subfloat{
    \includegraphics[width=0.225\textwidth]{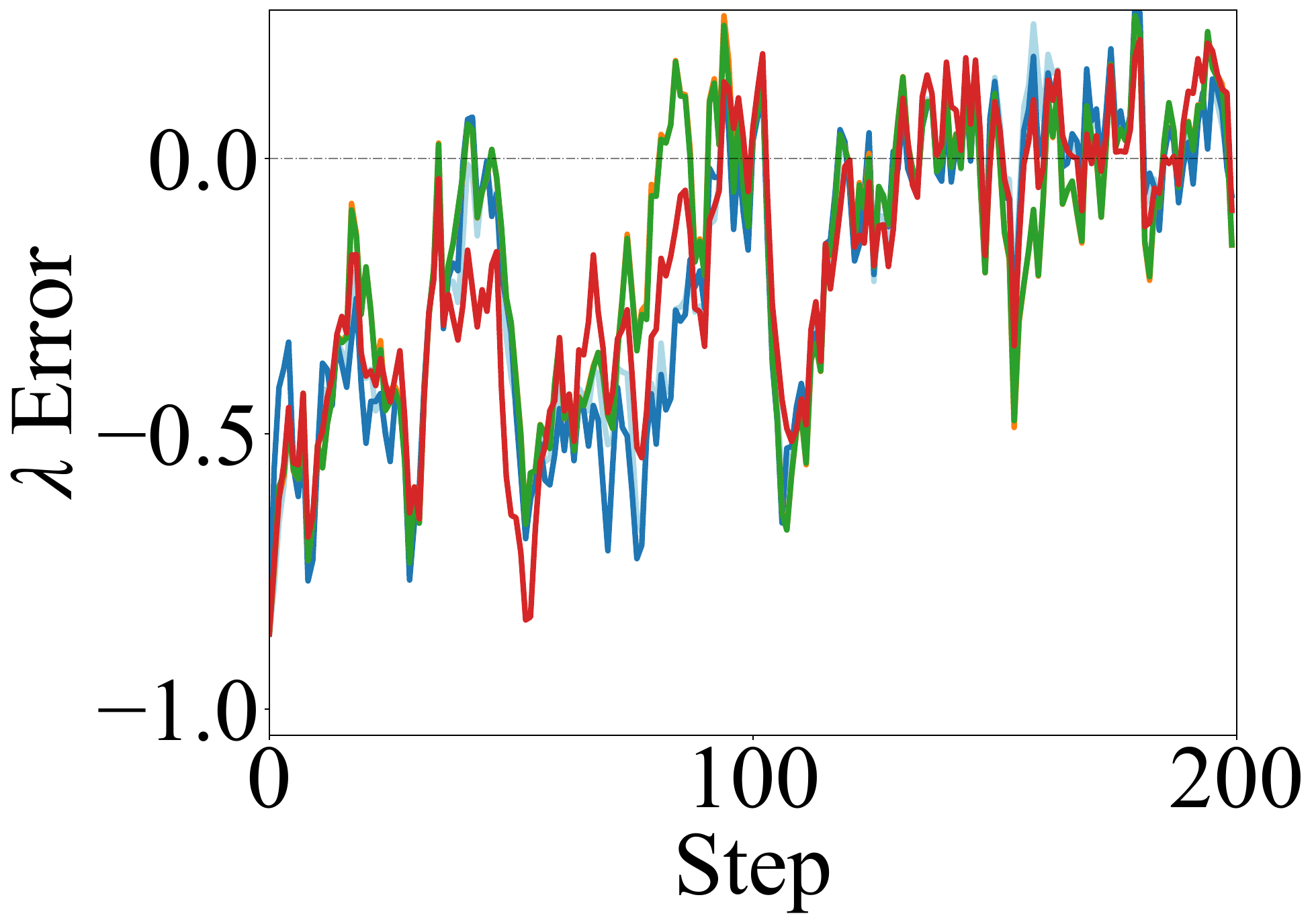}
    \hfill
    \includegraphics[width=0.225\textwidth]{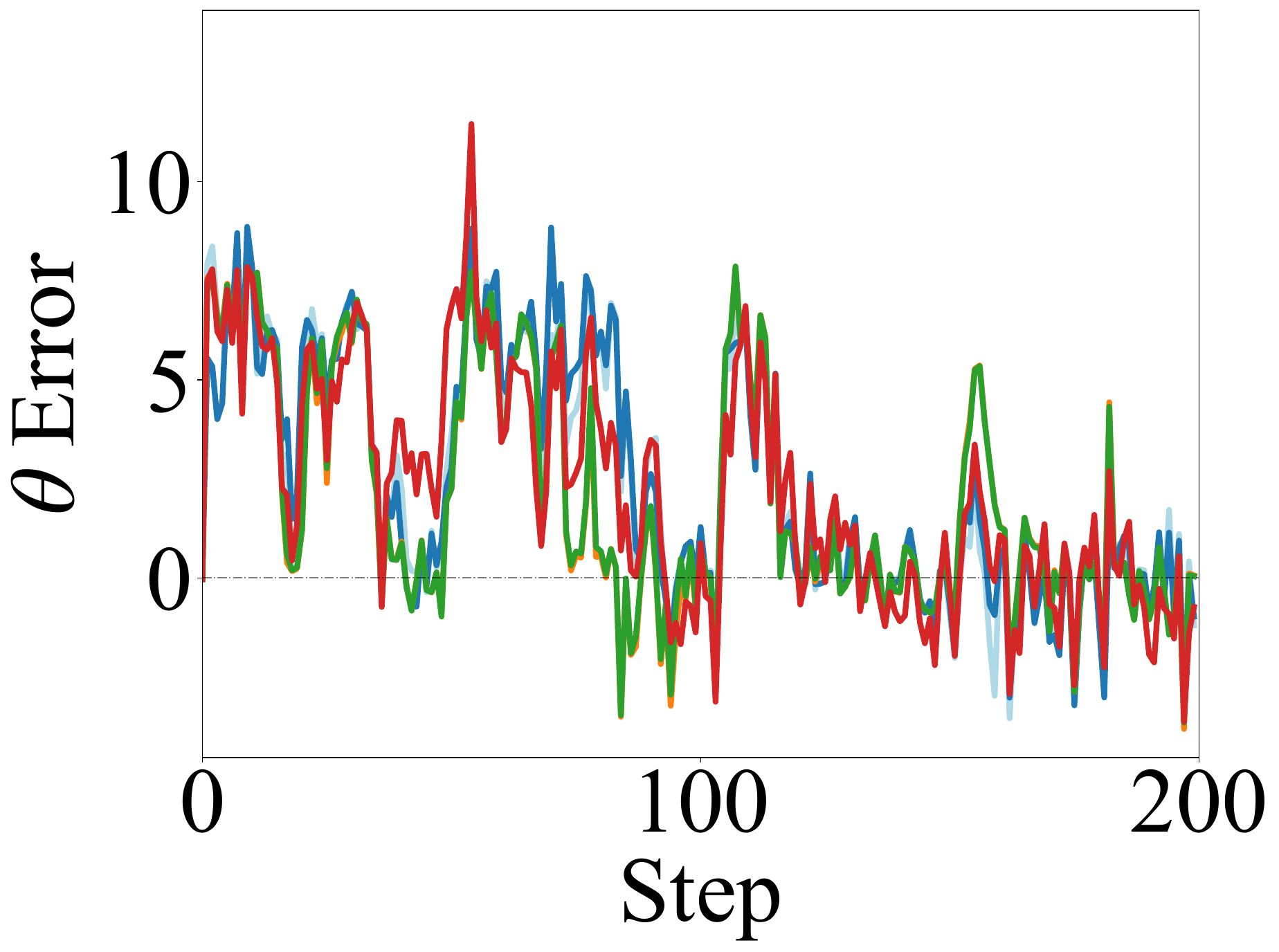}
    }
    \\
    \subfloat{
        \includegraphics[width=0.4\textwidth]{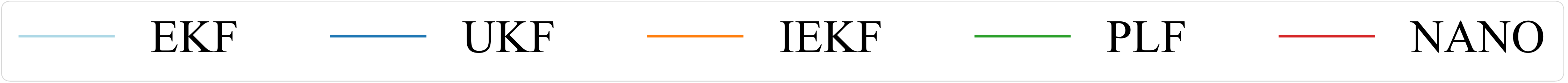}
        \phantomsection
    }
    \caption{ Estimation errors in FM Demodulator system.}
    \label{fig.Demodulator_error}
\end{figure}

\begin{figure}[!t]
    \centering
    \subfloat{
    \includegraphics[width=0.23\textwidth]{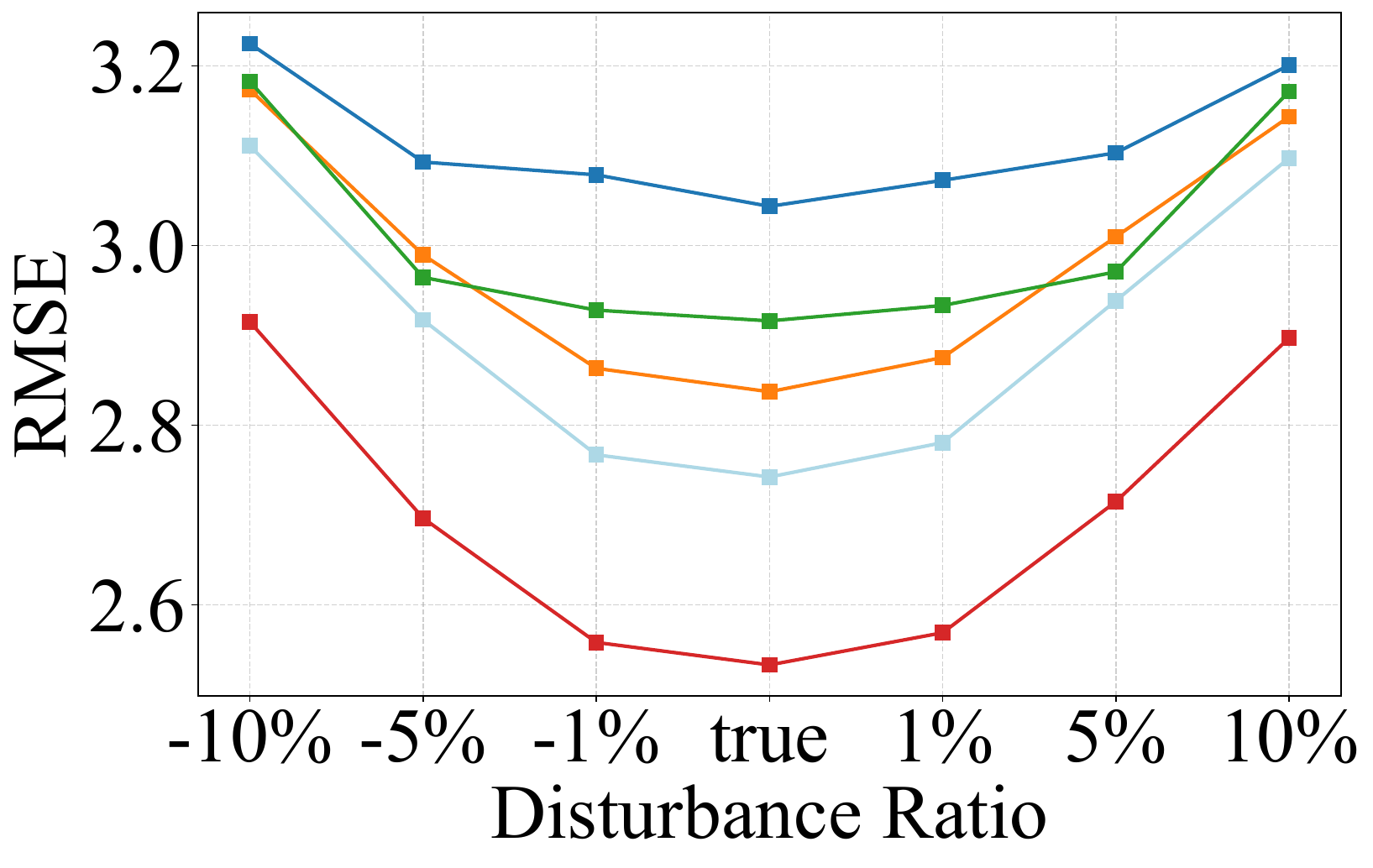}
    \hfill
    \includegraphics[width=0.22\textwidth]{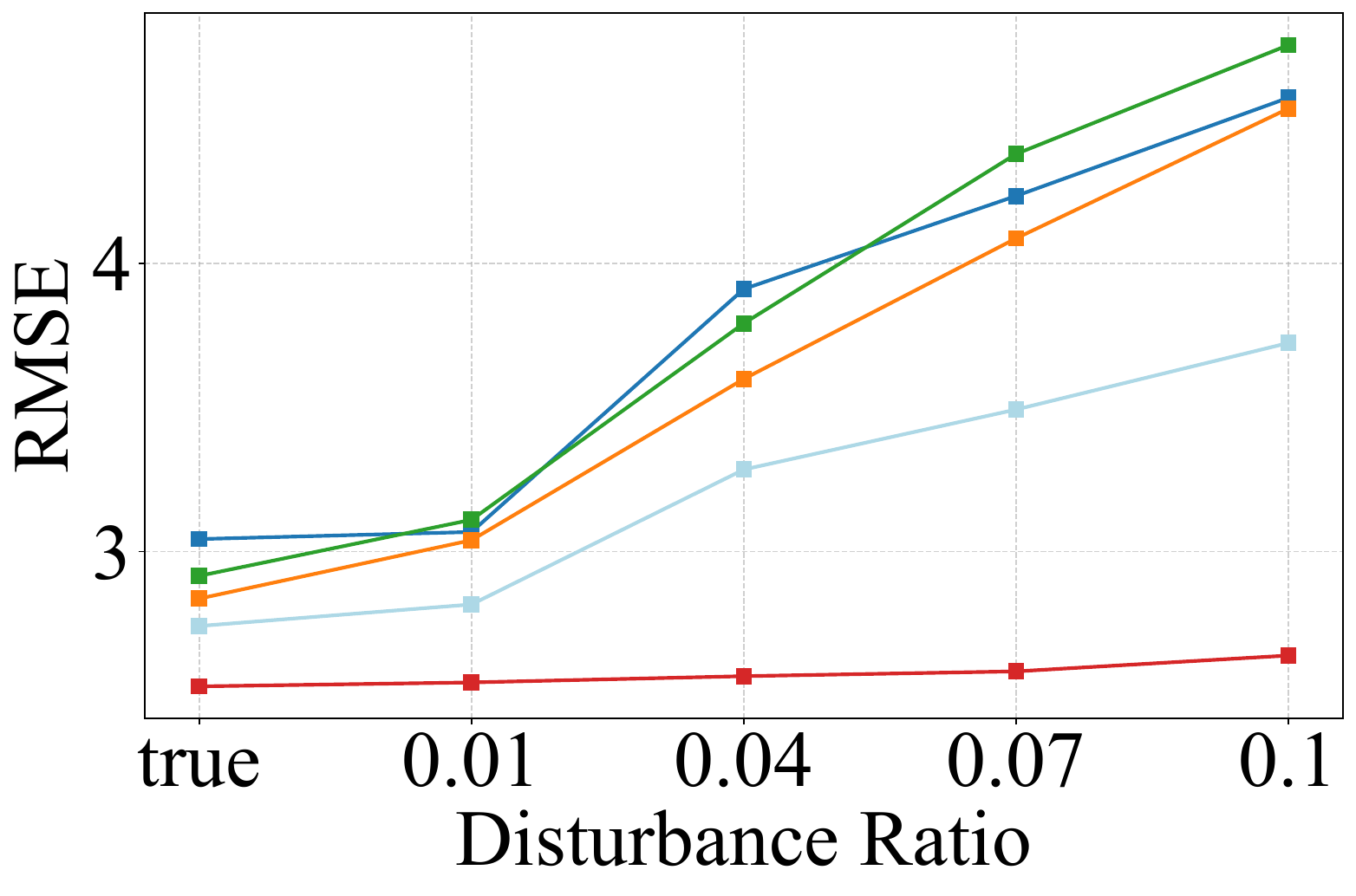}
    }
    \\
    \subfloat{
        \includegraphics[width=0.4\textwidth]{figures/legend.pdf}
        \phantomsection
    }
    \caption{Mean RMSE of FM Demodulator system under model mismatch condition. Left: system modeling error, Right: measurement outlier.}
    \label{fig.Demodulator_outlier}
\end{figure}

Fig.~\ref{fig.Demodulator} and Fig.~\ref{fig.Demodulator_error} show the experimental results under ideal conditions, where NANO has the lowest error and its error curve is closest to zero. As shown in Fig.~\ref{fig.Demodulator_outlier}, when the system modeling error increases, the RMSE of all filters increases, but the RMSE of NANO remains the lowest. When the measurement outlier increases, the RMSE of NANO grows the slowest, demonstrating excellent robustness.

\subsection{Satellite Attitude Estimation}
Next, we consider a satellite attitude estimation problem where the attitude of the satellite is represented using Euler angles \cite{brossard2020code}. The state variable is defined as \(\boldsymbol{\theta} = \left[\theta_p, \theta_r, \theta_y\right]^\top \in \mathbb{R}^3\), where \(\theta_p, \theta_r, \theta_y\) correspond to the pitch, roll, and yaw angles, respectively. The control input is the angular velocity of the satellite, denoted as \(\omega\in\mathbb{R}^3\). The state space model of this system is 
\begin{equation}\nonumber
\begin{aligned}
\boldsymbol{\theta}_{t+1} &= \boldsymbol{\theta}_{t} + \Omega(\boldsymbol{\theta}_t)\omega_t\Delta t+ \xi_t, \\
y_t &= \begin{bmatrix}
    C(\boldsymbol{\theta}_t)^\top g \\
    C(\boldsymbol{\theta}_t)^\top b 
\end{bmatrix} + \zeta_t,
\end{aligned}
\end{equation}
with \(\Delta t = 0.01s\). The control input is set to  \(\boldsymbol{\omega}_t = \frac{\pi}{18}\sin(2\Delta t\pi t)\cdot \mathbf{1}_{3 \times 1}\), and the vectors \(g=\left[0, 0,-9.81\right]^\top\), \(b=\left[27.75, -3.65, 47.21\right]^\top\) represent the gravitational acceleration and Earth's magnetic field, respectively. The explicit forms of the transformation matrices \(\Omega(\theta)\) and \(C(\theta)\) can be found in \cite{brossard2020code}.
% The transformation matrices \(\Omega(\theta)\) and \(C(\theta)\) are given by 
% \begin{equation}\nonumber
% \begin{aligned}
% \Omega(\boldsymbol{\theta})  &=\left[\begin{array}{ccc}
% 1 & \left(s_p s_r\right) / c_p & \left(c_r s_p\right) / c_p \\
% 0 & c_r & -s_r \\
% 0 & s_r / c_p & c_r / c_p
% \end{array}\right],  \\
% C(\boldsymbol{\theta})&=\begin{bmatrix}
% c_y c_p & c_y s_p s_r - s_y c_r & c_y s_p c_r + s_y s_r \\
% s_y c_p & s_y s_p s_r + c_y c_r & s_y s_p c_r - c_y s_r \\
% -s_p & c_p s_r & c_p c_r
% \end{bmatrix},
% \end{aligned}
% \end{equation}
% where \(s_i = \sin(\theta_i)\), \( c_i=\cos(\theta_i)\). 
The initial state is set to \( x_0 \sim \mathcal{N}(0, 10^{-3}\cdot\mathbf{I}_{3 \times 3}) \). In this experiment, we consider a scenario where the process noise $\xi_t$ follows a Laplace distribution, and the measurement noise $\zeta_t$ follows Gaussian noise, as
\begin{equation*}
\begin{aligned}  
\xi_t \sim \text{Laplace}(0, 10^{-5} \cdot\mathbf{I}_{3 \times 3}),\quad \zeta_t \sim \mathcal{N}(0,  10^{-4} \cdot\mathbf{I}_{3 \times 3}).    
\end{aligned}
\end{equation*}
We also take into account two types of model mismatch: system modeling errors and measurement outliers:
% Note that in this setting, where outliers are present in the system, the ground truth in our simulations is generated using the noise model with outliers, $\xi_t$. However, the $Q$ and $R$ matrices for the filters are still configured based on the noise distribution without outliers.

\begin{itemize}
    \item System modeling error: \(\Delta t \leftarrow \Delta t (1 + o), o\in \{0, \pm10\%, \pm20\%,\pm30\%\}\)
    \item Measurement outlier: \(\zeta_t \sim (1-k)\mathcal{N}(0,  10^{-4} \cdot\mathbf{I}_{3 \times 3})+ k\text{Beta}(1.2, 1.5)\cdot \mathbf{1}_{3\times 1}\), \(k\in \{0, 0.01, 0.05,0.1,0.15\}\)
\end{itemize}

\begin{figure}[!t]
\centering
    \includegraphics[width=0.4\textwidth]{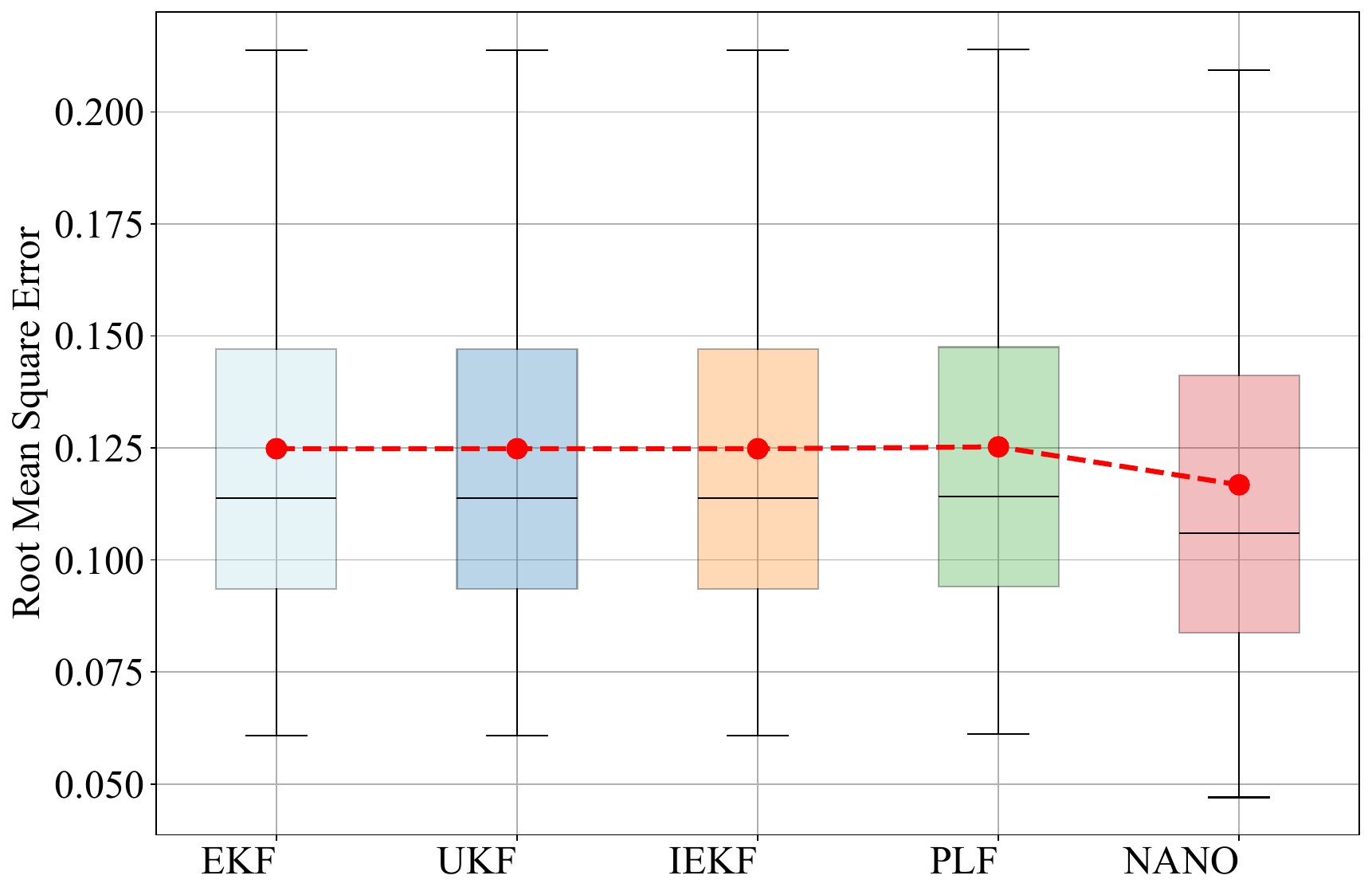}
    \label{fig.Attitude_gauss}

\caption{Box plot of RMSE over all MC experiments for satellite attitude estimation. Note that red point `` $\textcolor{red}{\bullet}$
 " represents the average RMSE.}
\label{fig.attitude}
\end{figure}

\begin{figure}[!t]
    \centering
    
    \includegraphics[width=0.157\textwidth]{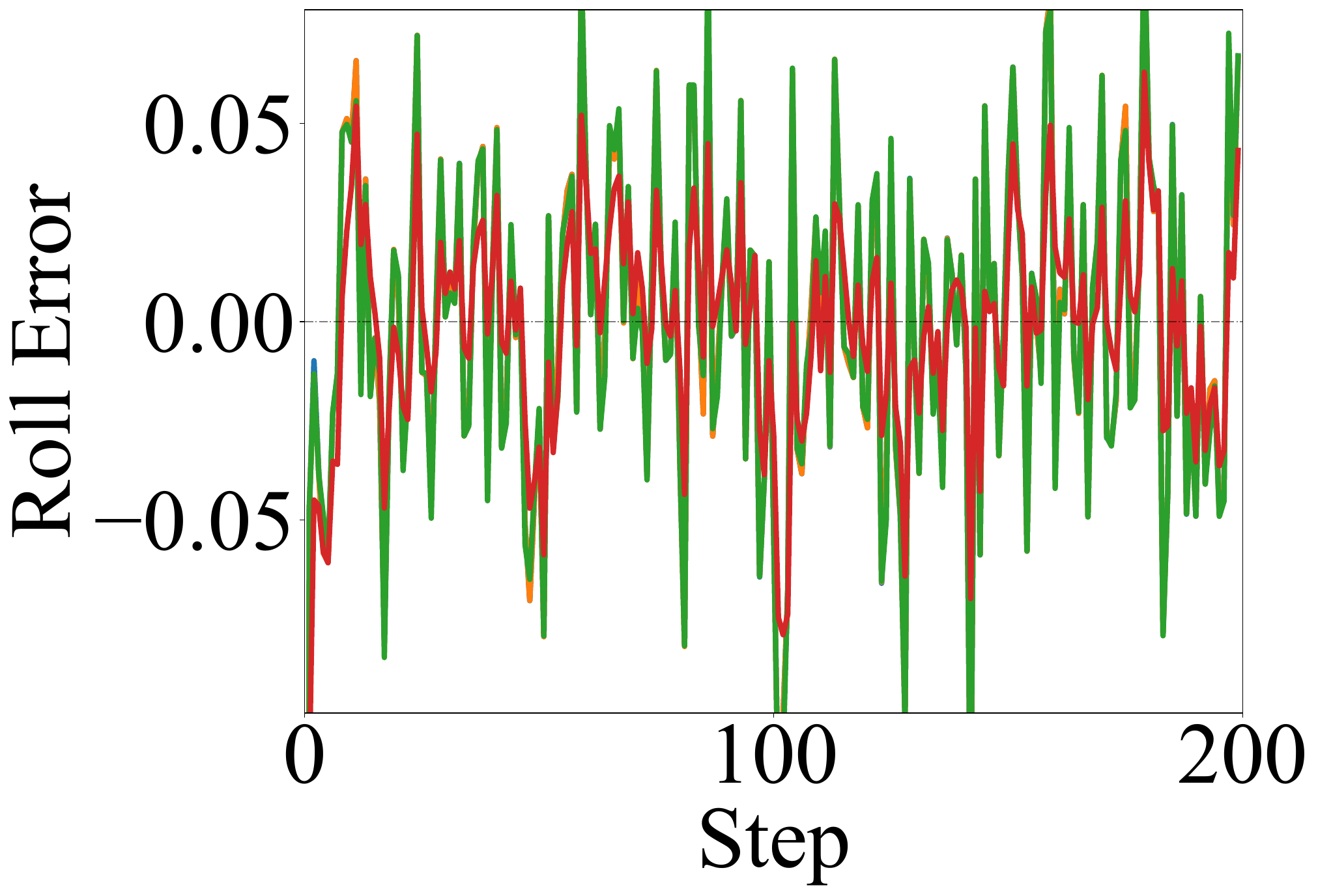}
    \hfill
    \includegraphics[width=0.157\textwidth]{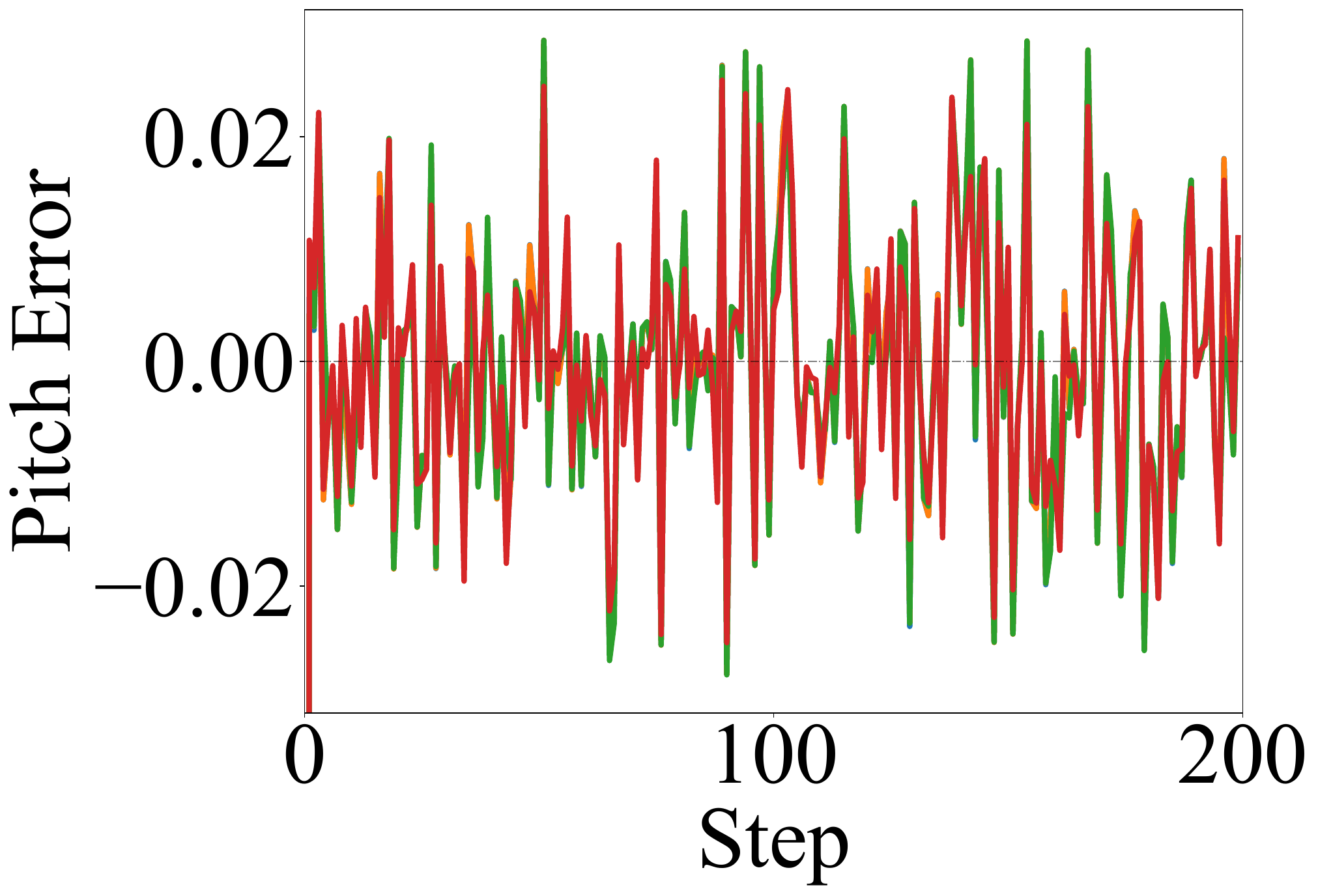}
    \hfill
    \includegraphics[width=0.157\textwidth]{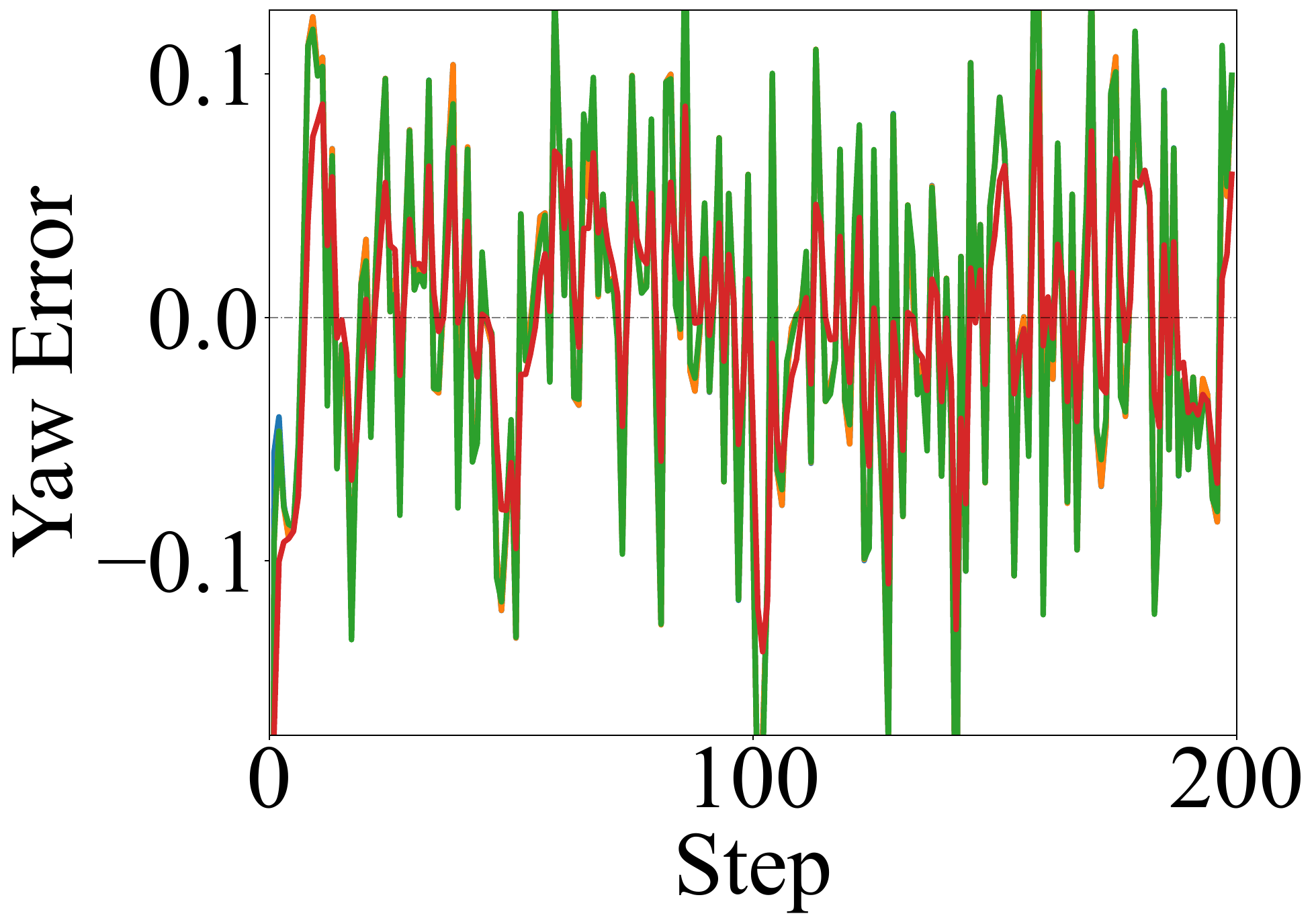}
    \\
    \subfloat{
        \includegraphics[width=0.4\textwidth]{figures/legend.pdf}
    }
    \caption{Estimation errors in satellite attitude estimation.}
    \label{fig.attitude_error}
\end{figure}

\begin{figure}[!t]
    \centering
    \subfloat{
    \includegraphics[width=0.23\textwidth]{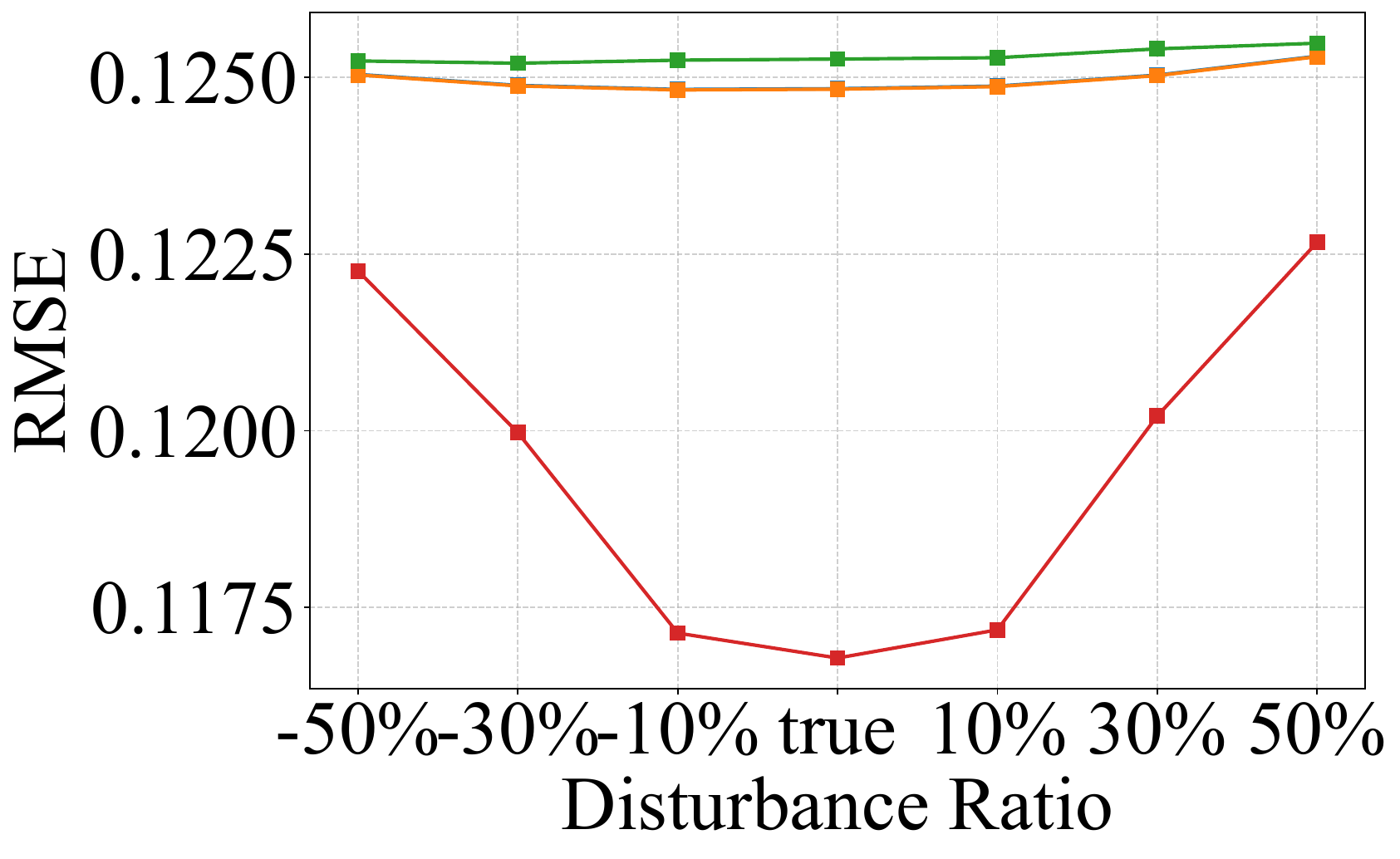}
    \hfill
    \includegraphics[width=0.22\textwidth]{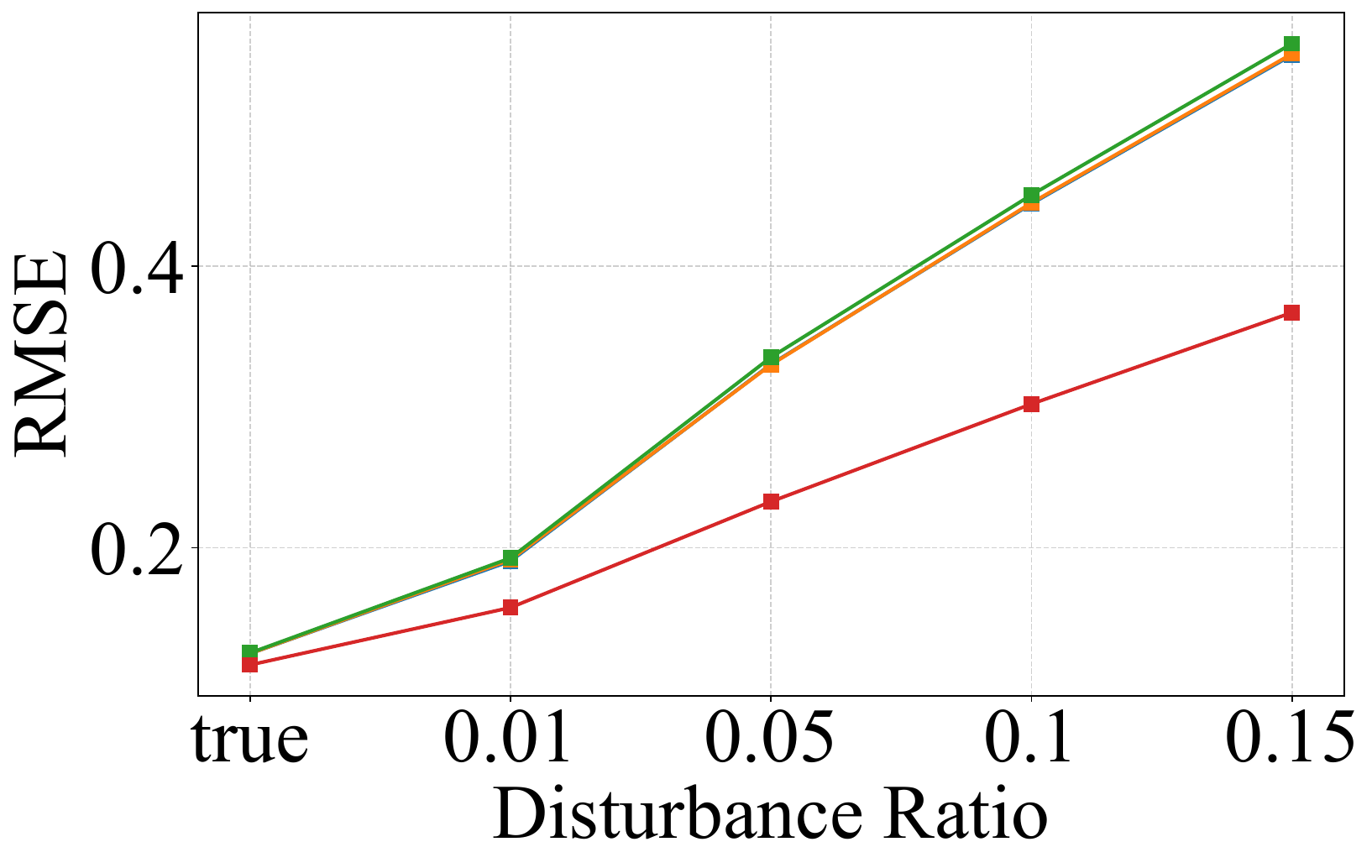}
    }
    \\
    \subfloat{
        \includegraphics[width=0.4\textwidth]{figures/legend.pdf}
        \phantomsection
    }
    \caption{Mean RMSE of satellite attitude system under model mismatch condition. Left: system modeling error, Right: measurement outlier.}
    \label{fig.attitude_outlier}
\end{figure}

Fig.~\ref{fig.attitude} and Fig.~\ref{fig.attitude_error} show that the NANO filter still outperforms all other Gaussian filters under ideal conditions. In Fig.~\ref{fig.attitude_outlier}, as the system modeling error increases, the estimation error of the NANO filter increases, but it still remains smaller than that of other methods all the time. Similar experimental results are also shown in the case of measurement outliers.

\subsection{Duffing Oscillator}
Finally, we consider a Duffing oscillator system, a canonical nonlinear benchmark for state estimation studies \cite{guckenheimer2013nonlinear}. The state vector is the displacement–velocity pair \(\left[x_{t+1}, \dot x_{t+1}\right]\). The state space model of this system is
\begin{equation}
\begin{aligned}
\begin{bmatrix}
x_{t+1}\\
\dot x_{t+1}
\end{bmatrix}
&=
\begin{bmatrix}
x_{t} + \dot x_{t}\,\Delta t\\[4pt]
\dot x_{t} \;+\;\bigl(-2\zeta\omega_{n}\dot x_{t} - \omega_{n}^{2}x_{t} - \beta x_{t}^{3}+ F\cos(\Omega t_{k})\bigr)\Delta t
\end{bmatrix}
+ \boldsymbol{w}_{t},\\[8pt]
y_{t} &= x_{t}^{3} + v_{t}.
\end{aligned}
\end{equation}
Here, $\Delta t = 0.01\,\text{s}$, $\zeta = 0.05$, $\omega_{n}=1\,\text{rad/s}$, $\beta = 1$, $F = 0.2$, $\Omega = 1.2\,\text{rad/s}$, $w_t \sim \mathcal{N}(\mathbf{0}, 10^{-3}\cdot\mathbf{I}_{2\times2})$, and $v_t \sim \mathcal{N}(0, 10^{-2})$. We also take into account two types of model mismatch: system modeling errors and measurement outliers:
\begin{itemize}
    \item System modeling error: \(\omega_n \leftarrow \omega_n (1 + o), o\in \{0, \pm10\%, \pm20\%,\pm30\%\}\)
    \item Measurement outlier: \(\zeta_t \sim (1-k)\mathcal{N}(0,  10^{-2})+ k\mathcal{N}(0, 1)\), \(k\in \{0, 0.03, 0.05,0.08,0.1\}\)
\end{itemize}

\begin{figure}[!t]
\centering
\includegraphics[width=0.4\textwidth]{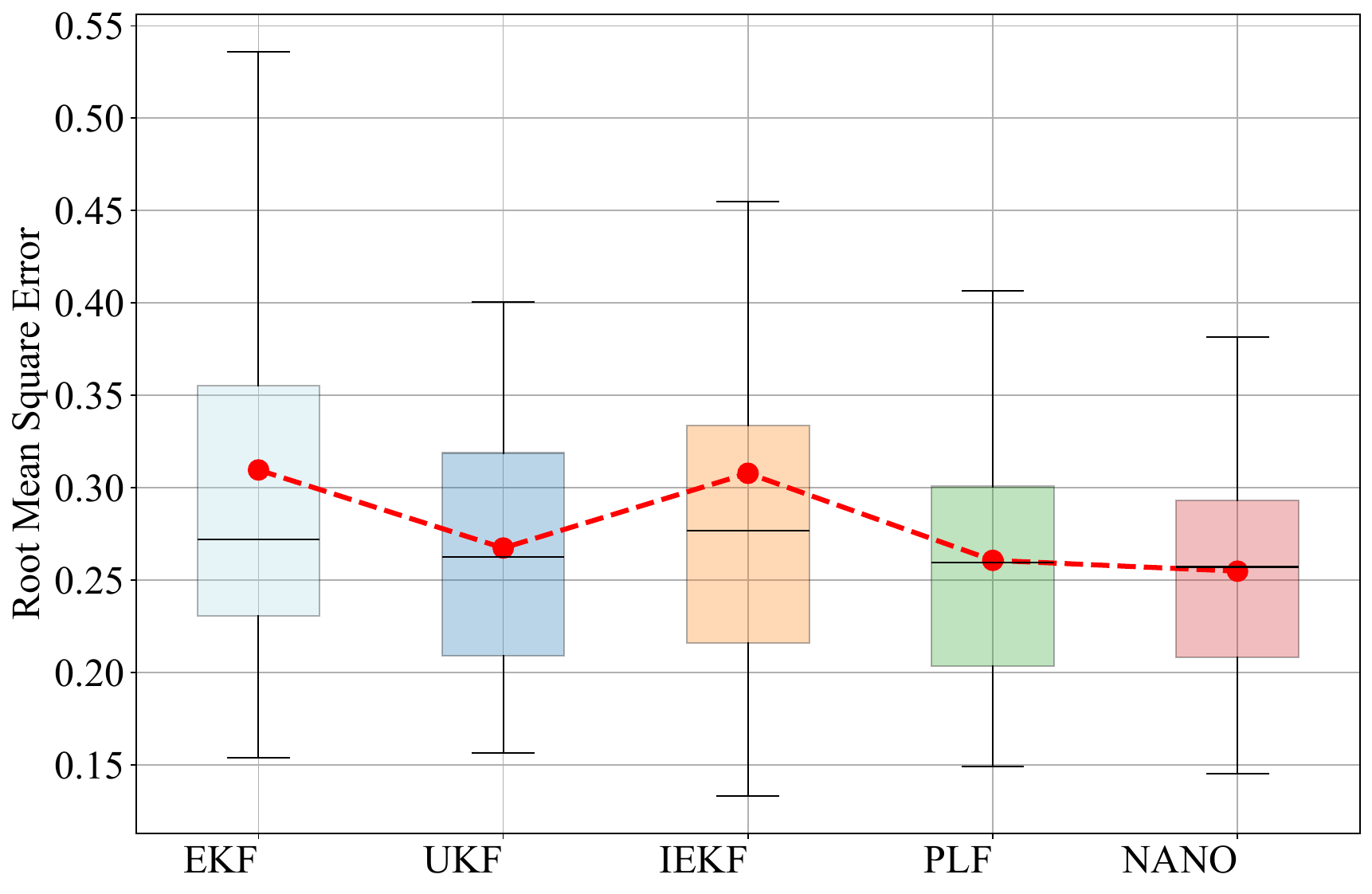}

\caption{Box plot of RMSE over all MC experiments for Duffing Oscillator. Note that red point `` $\textcolor{red}{\bullet}$
 " represents the average RMSE.}
\label{fig.Duffing}
\end{figure}

\begin{figure}[!t]
    \centering
    \subfloat{
    \includegraphics[width=0.225\textwidth]{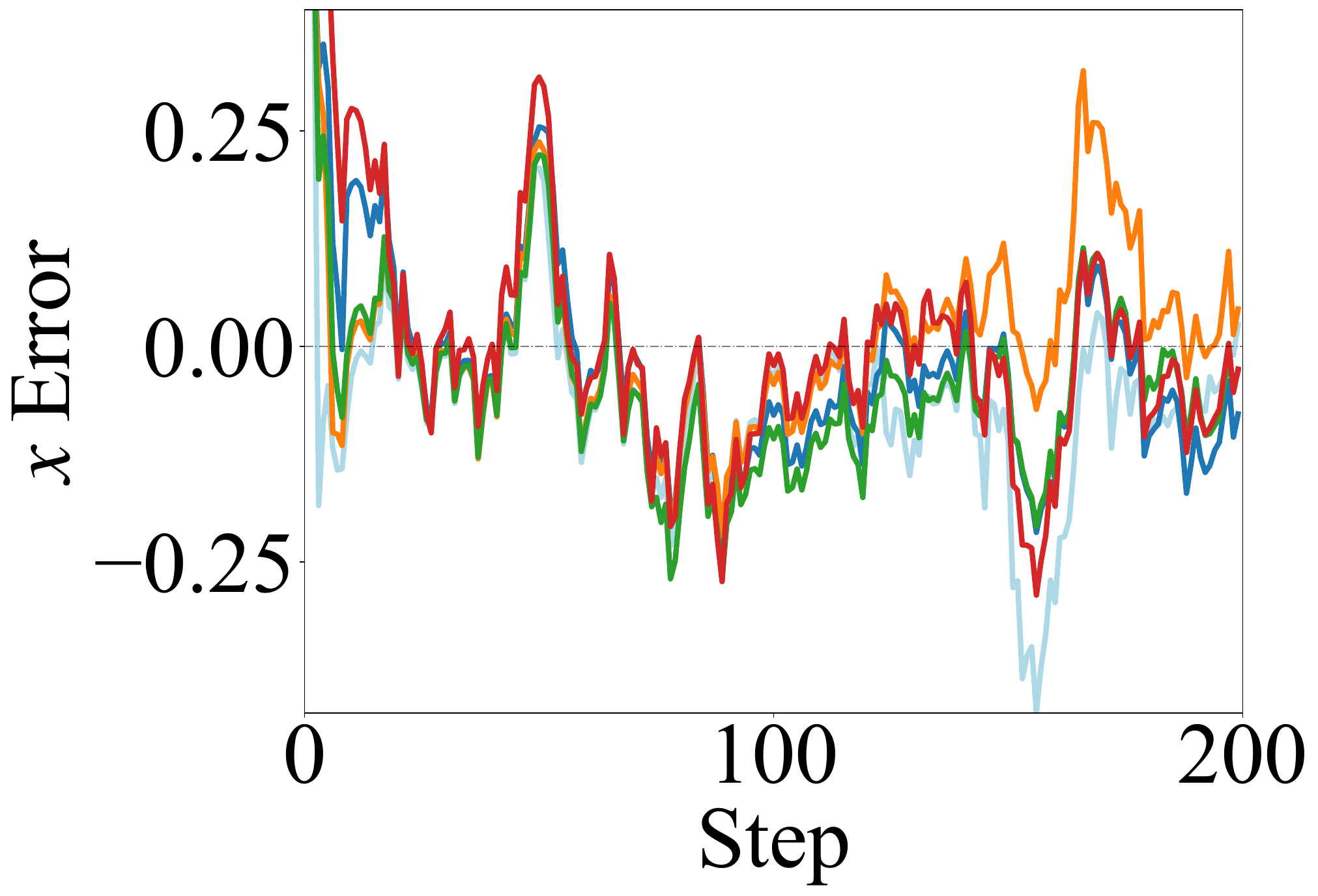}
    \hfill
    \includegraphics[width=0.225\textwidth]{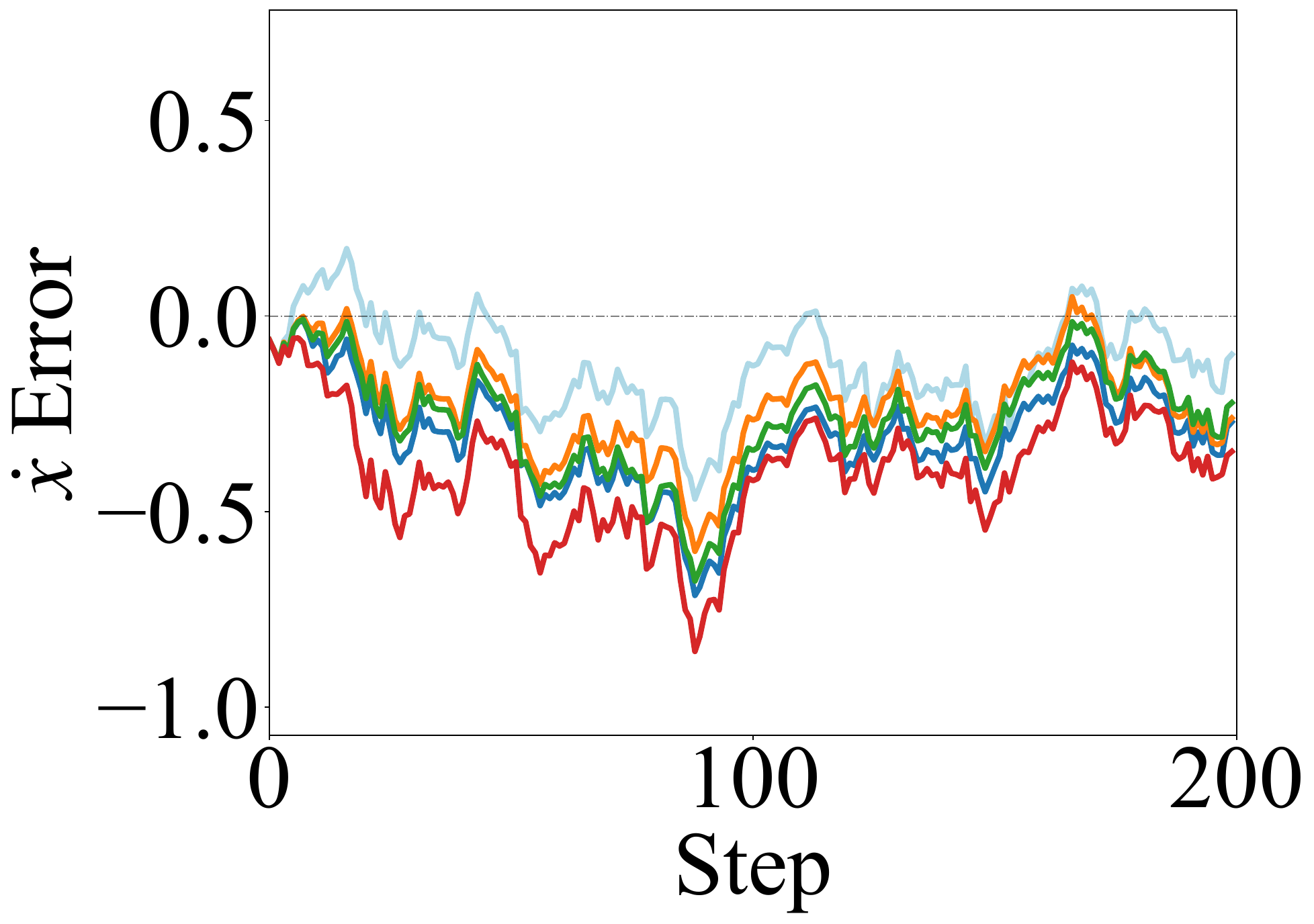}
    }
    \\
    \subfloat{
        \includegraphics[width=0.4\textwidth]{figures/legend.pdf}
        \phantomsection
    }
    \caption{Estimation errors in Duffing Oscillator system.}
    \label{fig.Duffing_error}
\end{figure}

\begin{figure}[!t]
    \centering
    \subfloat{
    \includegraphics[width=0.225\textwidth]{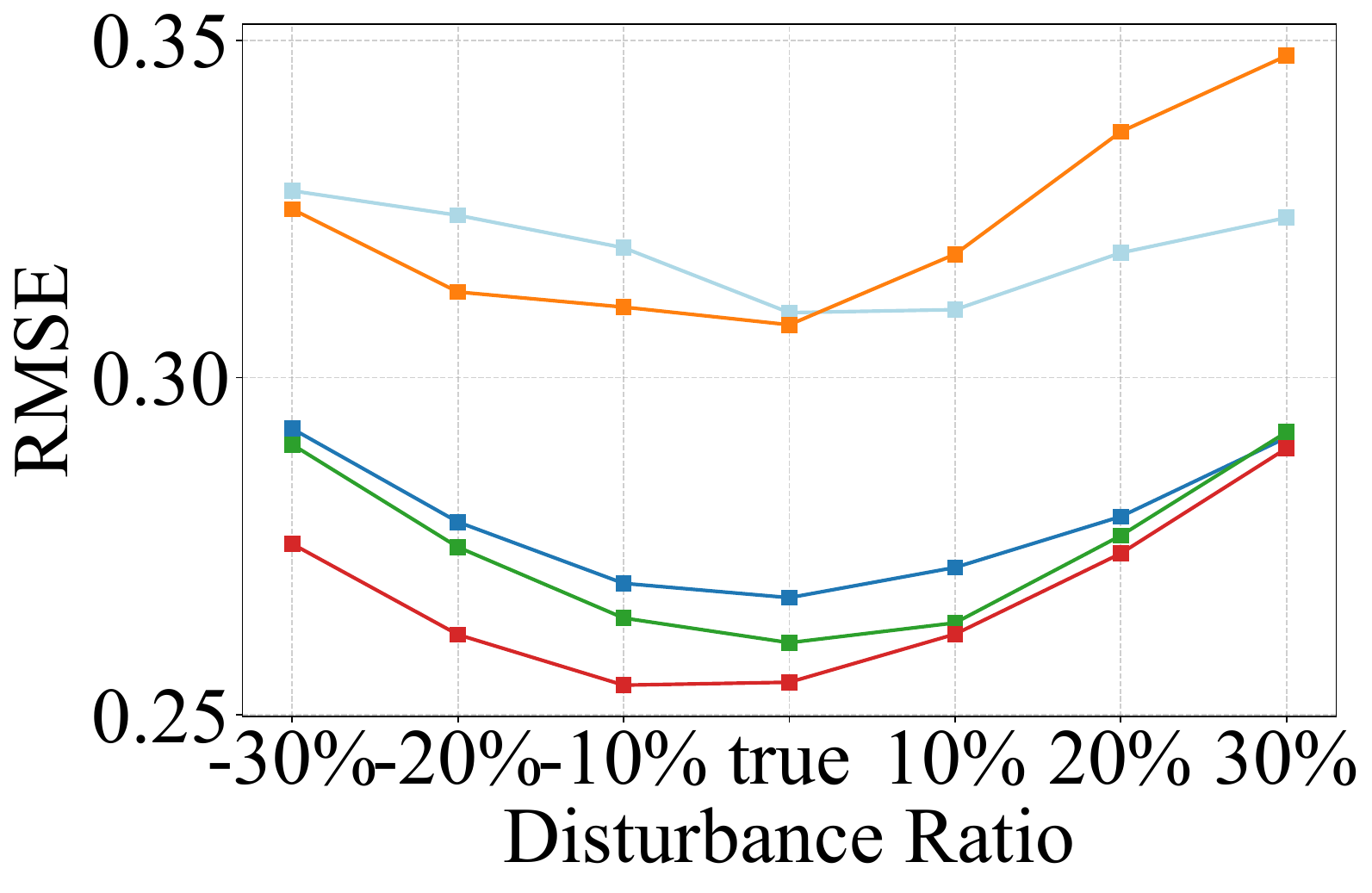}
    \hfill
    \includegraphics[width=0.225\textwidth]{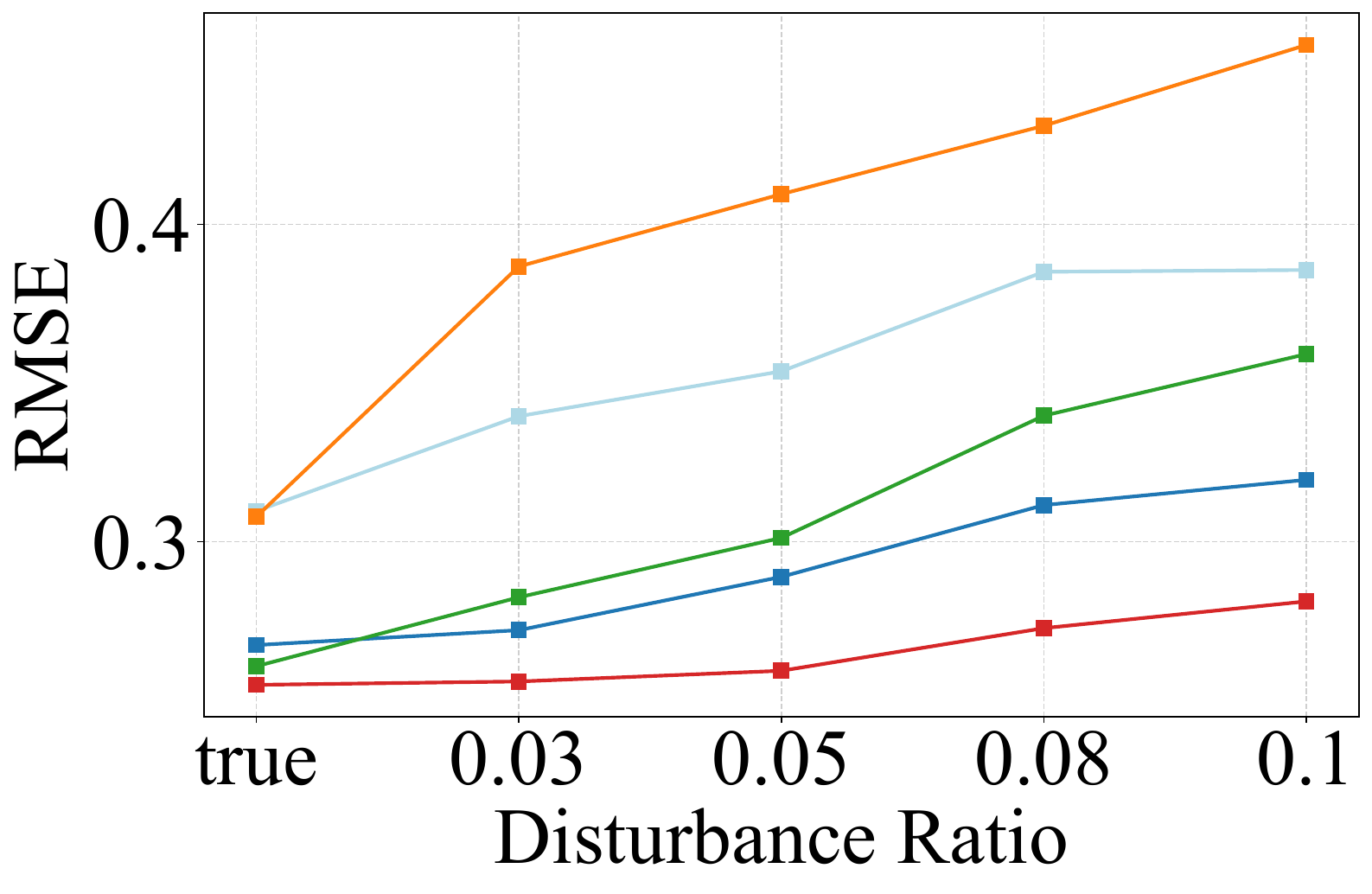}
    }
    \\
    \subfloat{
        \includegraphics[width=0.4\textwidth]{figures/legend.pdf}
        \phantomsection
    }
    \caption{Mean RMSE of Duffing Oscillator system under model mismatch condition. Left: system modeling error, Right: measurement outlier.}
    \label{fig.Duffing_outlier}
\end{figure}

As shown in Fig.~\ref{fig.Duffing} and Fig.~\ref{fig.Duffing_error}, the NANO filter achieves a lower average RMSE compared to the other filters, due to its superior effectiveness in handling nonlinear dynamics. It also indicates that the performance of NANO does not degrade after incorporating our positive definiteness guarantee. 
Fig.~\ref{fig.Duffing_outlier} also shows that the NANO filter can achieve the best results when there is a model mismatch.

The three experiments above demonstrate the effectiveness of our proposed NANO filter with guaranteed positive definiteness compared to traditional members of the Kalman filter family. Finally, we conduct an ablation study to compare our method with two baselines: the original derivative-based NANO without positive definiteness guarantee (NANO w/o P) and its EKF-initialized variant (NANO-EKF). The results, summarized in Table~\ref{table:results}, show that our NANO with positive definiteness guarantee consistently achieves the lowest RMSE across all three environments. In contrast, NANO w/o P diverges in the FM demodulator scenario due to the loss of positive definiteness in the covariance matrix, while NANO-EKF performs poorly because EKF provides only suboptimal initialization points in highly nonlinear systems. Moreover, since our method avoids the computation of the Hessian matrix, it also achieves the higher computational efficiency compared to the original NANO filter.

\begin{table}[t]
\scriptsize % 设置表格字体大小
\caption{{Ablation studies of NANO on different environments.}} 
\label{table:results}
\centering
\resizebox{0.5\textwidth}{!}{
\begin{tabular}{*{9}{c}}\toprule
\multirow{1}{*}{} & \multicolumn{2}{c}{\multirow{1}{*}{FM Demodulator}} 
& \multicolumn{2}{c}{\multirow{1}{*}{{Attitude Estimation}}}
& \multicolumn{2}{c}{\multirow{1}{*}{{Duffing Oscillator}}}
\\ \midrule
{} & 
\multicolumn{1}{c}{{RMSE}} &
\multicolumn{1}{c}{{time (ms)}} &
\multicolumn{1}{c}{{RMSE}} & 
\multicolumn{1}{c}{{time (ms)}} &
\multicolumn{1}{c}{{RMSE}} &
\multicolumn{1}{c}{{time (ms)}} \\ 
\midrule
\multirow{1}{*}{NANO w/o P} & 
\multirow{1}{*}{diverge}  & \multirow{1}{*}{1.218} & \multirow{1}{*}{0.161} & 
\multirow{1}{*}{6.137} & \multirow{1}{*}{0.293} &   \multirow{1}{*}{0.819} \\ 

\multirow{1}{*}{{NANO-EKF}} & 
\multirow{1}{*}{2.829} & \multirow{1}{*}{1.426} & \multirow{1}{*}{0.127} & 
\multirow{1}{*}{6.137} &  \multirow{1}{*}{0.263} &  \multirow{1}{*}{0.922} \\

\multirow{1}{*}{NANO} & 
\multirow{1}{*}{\textbf{2.533}}  & \multirow{1}{*}{\textbf{0.481}} & \multirow{1}{*}{\textbf{0.116}} & 
\multirow{1}{*}{\textbf{1.868}} & \multirow{1}{*}{\textbf{0.254}} &   \multirow{1}{*}{\textbf{0.248}} \\ 
\bottomrule
\end{tabular}}
\end{table}

\section{Conclusion}
This paper identifies that the loss of positive definiteness in the covariance update of the NANO filter originates from the indefiniteness of the log-likelihood Hessian. To address this issue, we propose two methods to guarantee positive definiteness: the first applies the Gauss–Newton method, reformulating the Hessian as the self-adjoint product of the Jacobian of the normalized measurement residual; the second performs Cholesky decomposition of the covariance and iteratively updates the decomposed matrix, reconstructing the covariance through its Gram matrix representation. In our experiments, we adopt the Hessian approximation method. Comparative studies on various nonlinear systems demonstrate that the proposed NANO filter with positive definiteness guarantee consistently achieves the lowest estimation error compared with both the Kalman filter family and the original NANO filter.

\bibliographystyle{IEEEtran}
\bibliography{ref}

\end{document}